\documentstyle[12pt]{article}

\title {Multivariate hypergeometric functions as tau functions
of Toda lattice and Kadomtsev-Petviashvili equation}

\author{A.Yu.\ Orlov \\
\emph{e-mail: orlovs@kusm.kyoto-u.ac.jp and orlovs@wave.sio.rssi.ru}\\
\emph{Department of Mathematics, Faculty of Science, Kyoto University}\\
\emph{\ Kyoto 606-85-02, Japan.}\\
\emph{permanent address: Nonlinear wave processes laboratory}\\
\emph{Oceanology Institute, 36 Nakhimovskii prospekt,}\\
\emph{Moscow 117851, Russia.} \and D.M.\ Scherbin \\
\emph{e-mail: scherbin@wave.sio.rssi.ru}\\
\emph{Nonlinear wave processes laboratory}\\
\emph{Oceanology Institute, 36 Nakhimovskii prospekt,}\\
\emph{Moscow 117851, Russia.}}

\small  \oddsidemargin -10mm  \evensidemargin 5mm  \topmargin -10mm
\def\be{\begin{equation}}
\def\ee{\end{equation}}
\def\g{\gamma}

\renewcommand{\l}{\langle}
\renewcommand{\r}{\rangle}

\begin{document}
\maketitle

\begin{abstract}
We present the q-deformed multivariate hypergeometric
functions related to Schur polynomials as tau-functions
of the KP and of the two-dimensional 
Toda lattice hierarchies. The variables of the hypergeometric functions
are the higher times of those hierarchies. The discrete Toda lattice 
variable shifts parameters of hypergeometric functions. 
The role of additional symmetries in generating hypergeometric
tau-functions is explained.
\end{abstract}

\section{Introduction}

Hypergeometric functions play an important role both in physics and in
mathematics \cite{GGR}. Many special functions and polynomials 
(such as q-Askey-Wilson polynomials, q-Jacobi polynomials, q-Racah polynomials, q-Hahn polynomials, expressions for Clebsch-Gordan coefficients) are just
certain hypergeometric functions evaluated in special values of parameters.
In physics hypergeometric functions and their q-deformed counterparts 
sometimes play the role of wave functions and
correlation functions for quantum integrable systems.
In the present paper we shall construct hypergeometric functions as
 tau-functions of the Kadomtsev-Petviashvili (KP) hierarchy of 
equations. 
It is interesting that the KP equation, which originally serves in plasma physics, now plays a very important role in physics \cite{ZM} (for modern applications see review \cite{M} ) and in mathematics. This peculiarity of Kadomtsev-Petviashvili
equation appeared in the paper \cite{Dr}, where L-A pair of the KP
equation was presented, and mainly in the paper of V.E.Zakharov and 
A.B.Shabat in 1974 where this equation was integrated by the dressing method. Actually it was the paper \cite{ZSh} where so-called hierarchy of higher KP 
equations appeared.
 Another very important equation is the two-dimensional Toda lattice
integrated first in \cite{AM}. In the present paper we use these equations to construct hypergeometric functions which depend on many variables, these variables are KP and Toda lattice higher times. Here we shall use the general approach to integrable hierarchies of Kyoto school \cite{JM}. Especially a set
of papers about Toda lattice \cite{UT,Tinit,TT,NTT,T} is important for us.\\
We briefly outline the connection of what we do with Zakharov-Shabat
dressing method \cite{ZSh} and with the nonlocal $\bar{\partial}$ problem \cite{ZMan}; we mention the related system of orthogonal polynomials.
The lack of space does not allow us to develop these 
topics.\\
We devote this paper to Vladimir Evgen'evich Zakharov on his 60 birthday.

\section{Notations}

There are several well-known different multivariate generalizations of
hypergeometric series of one variable \cite{V,KV}. 
Let $|q|<1$ and let ${\bf x}_{(N)}=(x_1,\cdots,x_N)$ be indeterminates. 
The multiple basic hypergeometric series \cite{Milne, KV} is
\begin{eqnarray}\label{vh}
{}_p\Phi_s\left.\left(a_1,\dots ,a_p\atop b_1,\dots ,b_s\right|q,{\bf x}_{(N)}\right)
=\sum_{l({\bf n})\le N}\frac {(q^{a_1};q)_{{\bf n}}\cdots (q^{a_p};q)_{{\bf n}}}
                            {(q^{b_1};q)_{{\bf n}}\cdots (q^{b_s};q)_{{\bf n}}}
\frac{q^{n({\bf n})}}{H_{{\bf n}}(q)}
s_{{\bf n}} \left( {\bf x}_{(N)} \right).
\end{eqnarray}
The sum is over all different partitions 
${\bf n}=\left(n_1,n_2,\dots,n_r\right)$, where 
$n_1\ge n_2\ge \cdots \ge n_r$, $r\le |{\bf n}|$, 
$|{\bf n}|=n_1+\cdots+n_r$ and whose length $l({\bf n})=r\le N$. 
Schur polynomial  $s_{{\bf n}} \left( {\bf x}_{(N)} \right)$, with 
$N \ge l({\bf n})$, 
is a symmetric function of variables
${\bf x}_{(N)}$ and  defined as follows \cite{Mac}:
\be\label{Schur}
s_{{\bf n}}({\bf x}_{(N)})=\frac{a_{{\bf n}+\delta}}{a_{\delta}},\quad
a_{{\bf n}}=\det (x_i^{n_j})_{1\le i,j\le N},\quad \delta=
(N-1, N-2,\dots,1,0).
\ee
Schur function $s_{{\bf n}}({\bf x}_{(N)})=0$ for $N<l({\bf n})$. 
In the theory of the KP hierarchy \cite{JM} 
it is convenient to define Schur functions in terms of KP higher
times ${\bf t}=(t_1,t_2,\dots)$ as
\be\label{KPvar}
s_{{\bf n}}({\bf t})=\det(p_{n_i-i+j}({\bf t}))_{1\le i,j\le r},\quad
\sum_{m=0}^{+\infty}p_m({\bf t})z^m=
\exp (\sum_{i=1}^{+\infty}t_iz^i)=e^{\xi({\bf t},z)}.
\ee
The functions $s_{{\bf n}}({\bf t})$ are related
to $s_{{\bf n}} \left( {\bf x}_{(N)} \right)$ 
 via the Miwa's change of variables:
\be\label{tmiwa}
t_m=\sum_{i=1}^{N} \frac{x_i^m}{m}.
\ee  
Each coefficient $(q^a;q)_{\bf n}$ in (\ref{vh}) is
expressed in terms of the so-called q-deformed Pochhammer symbols 
$(q^a;q)_{n_i}$:
\begin{eqnarray}\label{PSimb}
(q^a;q)_{\bf n}=(q^a ;q)_{n_1}(q^{a-1};q)_{n_2}\cdots (q^{a-r+1};q)_{n_r}\\
(q^a;q)_{n_i}=(1-q^a)(1-q^{a+1})\cdots(1-q^{a+n_i-1}), \quad (q^a;q)_0=1
\end{eqnarray}
$q$-deformed  'hook polynomials' $H_{{\bf n}}(q)$ are 
\begin{eqnarray}\label{hp}
H_{{\bf n}}(q)=\prod_{(i,j)\in {\bf n}} \left(1-q^{h_{ij}}\right),\quad
h_{ij}=(n_i+n'_j-i-j+1),
\end{eqnarray}
where ${\bf n'}=\left({n'}_1+{n'}_2+\cdots+{n'}_{r'}\right)$ is the 
conjugated partition \cite{Mac} and 
$q^{n({\bf n})}=q^{\sum_{i=1}^r (i-1)n_i}$.\\

The formula
\[
{}_p\Phi_s\left.\left(a_1,\dots ,a_p\atop b_1,\dots ,b_s\right|q,
{\bf x}_{(N)},{\bf y}_{(N)}\right)=
\]
\be
\sum_{l({\bf n})\le N}\frac {(q^{a_1};q)_{{\bf n}}\cdots (q^{a_p};q)_{{\bf n}}}
                            {(q^{b_1};q)_{{\bf n}}\cdots (q^{b_s};q)_{{\bf n}}}
\frac{q^{n({\bf n})}}{H_{{\bf n}}(q)}
\frac
{s_{{\bf n}} \left( {\bf x}_{(N)} \right)s_{{\bf n}} \left( {\bf y}_{(N)} \right)}
{s_{{\bf n}} \left(1,q,q^2,...,q^{N-1} \right)} \label{vh2}
\ee
defines the multiple basic hypergeometric function of two sets of variables
\cite{KV}.
\\
Another generalization of hypergeometric series is so-called 
hypergeometric function of matrix argument ${\bf X}$ with 
indices $\bf{a}$ and $\bf{b}$ \cite{V}: 
\begin{eqnarray}\label{hZ}
{}_pF_s\left.\left(a_1,\dots ,a_p\atop b_1,\dots ,b_s\right|{{\bf{X}}}
\right)
=\sum_{{\bf n}}\frac {(a_1)_{{\bf n}} \cdots (a_p)_{{\bf n}}}
                     {(b_1)_{{\bf n}} \cdots (b_s)_{{\bf n}}}
\frac{Z_{{\bf n}}({\bf X})}
{|{\bf n}|!} .
\end{eqnarray}
Here ${\bf X}$ is $N\times N$ matrix, and $Z_{{\bf n}}({\bf X})$
is zonal spherical polynomial for the symmetric space $GL(N,C)/U(N)$, 
see \cite{V},\cite{KV}. 
Let us note that
in the limit $q \to 1$ series (\ref{vh}) 
coincides with (\ref{hZ}), see \cite{KV}.\\

Now let us review some facts from the KP theory \cite{JM}. 
We have fermionic fields $\psi(z)=\sum_{k\in Z} \psi_k z^k$ and $\psi^*(z)=\sum_{k\in Z} \psi^*_k z^{-k-1}dz$, where fermionic operators satisfy the canonical anti-commutation relations:
\be \label{antikom}
[\psi_m,\psi_n]_+=[\psi^*_m,\psi^*_n]_+=0;\qquad [\psi_m,\psi^*_n]_+=
\delta_{mn} .
\ee
Let us introduce left and right vacuums by the properties:
\begin{eqnarray}\label{vak}
\psi_m |0\r=0 \qquad (m<0),\qquad \psi_m^*|0\r =0 \qquad (m \ge 0), \\
\l 0|\psi_m=0 \qquad (m\ge 0),\qquad \l 0|\psi_m^*=0 \qquad (m<0).
\end{eqnarray}
Throughout the text the subscript $*$ does not denote the complex conjugation.
The vacuum expectation value is defined by relations:
\begin{eqnarray} 
\l 0|1|0 \r=1,\quad \l 0|\psi_m\psi_m^* |0\r=1\quad m<0, \quad  \l 0|\psi_m^*\psi_m |0\r=1\quad m\ge 0 ,
\end{eqnarray}
\be\label{end}
 \l 0|\psi_m\psi_n |0\r=\l 0|\psi^*_m\psi^*_n |0\r=0,\quad \l 0|\psi_m\psi_n^* |0\r=0 \quad m\ne n .
\ee
Let us notice that if a function $h$ has no poles and zeroes for 
integer value of argument, then relations (\ref{antikom})-(\ref{end}) 
are invariant under the following transformation
\begin{equation}\label{gauge}
\psi_n \mapsto \frac {1}{h(n)} \psi_n, 
\qquad \psi^*_n \mapsto h(n)\psi^*_n .
\end{equation}
Let us denote $\widehat{gl}(\infty)=
{\itshape{Lin}}\{ 1,:\psi_i\psi_j^*:|i,j \in Z\}$, 
with usual normal ordering 
$:\psi_i\psi_j^*:=\psi_i\psi_j^*-\l 0|\psi_i\psi_j^*|0\r$. 
We define the operator $g$ which is an element of the group  
$\widehat{GL}(\infty)$
corresponding to the infinite dimensional
Lie algebra $\widehat{gl}(\infty)$.
The tau-function of the KP equation and the tau-function 
of the two-dimensional Toda lattice (TL) sometimes are defined as
\be\label{taucorKP}
\tau_{KP}(M,{\bf t})=
\langle M|e^{H({\bf t})}g|M \rangle ,\quad M \in Z,
\ee
\be\label{taucor}
\tau_{TL}(M,{\bf t},{\bf t}^*)=
\langle M|e^{H({\bf t})}ge^{H^*({\bf t}^*)}|M \rangle ,\quad M \in Z,
\ee
where ${\bf t}=(t_1,t_2,\dots)$ and  ${\bf t}^*=(t_1^*,t_2^*,\dots)$ 
 are called higher Toda lattice times \cite{JM,UT} (the 
first set ${\bf t}$ is in the same time
the set of higher KP times). $H({\bf t})$ and $H^*({\bf t}^*)$ 
belong to the following ${\widehat {gl}}(\infty)$ Cartan subalgebras:
\be
H({\bf t})=\sum_{n=1}^{+\infty} t_n H_n ,\quad 
H^*({\bf t}^*)=\sum_{n=1}^{+\infty} t_n^* H_{-n},\quad 
H_n=\frac{1}{2\pi i}\oint :z^{n}\psi(z)\psi^*(z): .
\ee
According to \cite{JM} the integer $M$ in (\ref{taucor}) plays the role 
of discrete Toda lattice variable and defines the following charged vacuums
\begin{eqnarray}
\l M|=\l 0|\Psi^{*}_{M},\qquad |M\r=\Psi_{M}|0\r ,
\end{eqnarray}
\begin{eqnarray}
\Psi_{M}=\psi_{M-1}\cdots\psi_1\psi_0 \quad M>0,\qquad \Psi_{M}=\psi^{*}_{M}\cdots\psi^{*}_{-2}\psi^{*}_{-1}\quad M<0 , \nonumber\\
\Psi^{*}_{M}=\psi^{*}_{0}\psi^{*}_1\cdots\psi^{*}_{M-1} \quad M>0,\qquad \Psi^{*}_{M}=\psi_{-1}\psi_{-2}\cdots\psi_{M}\quad M<0 .
\end{eqnarray}

{\bf Lemma 1}\cite{JM} For $-j_1<\cdots <-j_r<0\le i_s<\cdots <i_1$, $s-r\ge 0$ the following formula is valid:
\be\label{glemma}
\l s-r|e^{H({\bf t})}\psi^*_{-j_1}\cdots
\psi^*_{-j_r}\psi_{i_s}\cdots\psi_{i_1}|0\r=
(-1)^{j_1+\dots +j_r+(r-s)(r-s+1)/2}s_{{\bf n}}({\bf t}) ,
\ee
where the partition ${\bf n}=(n_1,\dots , n_{s-r}, n_{s-r+1},\dots , n_{s-r+j_1})$ is defined by the following pair of partitions:
\begin{eqnarray}
(n_1, \dots , n_{s-r})=(i_1-(s-r)+1, i_2-(s-r)+2, \dots , i_{s-r}) ,\\
(n_{s-r+1},\dots , n_{s-r+j_1})=(i_{s-r+1},\dots , i_s|j_1-1,\dots , j_r-1).
\end{eqnarray}
Here $(\dots | \dots)$ is another notation for a partition \cite{Mac}. \\
So-called vertex operators $V(z)$ and $V^*(z)$ are defined by:
\be\label{vertex}
V(z)=z^Me^{\xi({\bf t},z)}e^{-\xi(\tilde{\partial},z^{-1})},\quad
V^*(z)=z^{-M}e^{-\xi({\bf t},z)}e^{\xi(\tilde{\partial},z^{-1})},
\ee
where $\tilde{\partial}=(\frac{\partial}{\partial t_1},
\frac{1}{2}\frac{\partial}
{\partial t_2},\dots)$ and the function $\xi({\bf t},z)$ is the same as 
in (\ref{KPvar}). 
The Baker-Akhiezer function and the conjugated one are
\be
w(M,{\bf t};z)=\frac{V(z)\tau}{\tau}
\qquad w^*(M,{\bf t};z)=\frac{V^*(z)\tau}{\tau} .
\ee

The function $u=2\partial_{t_1}^2\log \tau$ solves the celebrated KP equation
\be
u_{t_3}=\frac 14 u_{t_1t_1t_1}+\frac 32 uu_{t_1}+
\frac 34 \int^{t_1} u_{t_2t_2}dt_1' .
\ee

It is well-known that 
the Schur functions $s_{{\bf n}}({\bf t})$ are 
tau-functions for some rational solutions of the KP
hierarchy. It is also known that not any linear combination of Schur 
functions turns out to be a KP tau-function. In order to find these 
combinations one should solve bilinear difference equation (see \cite{JM}),
 a version of discrete Hirota equation. Below we shall present KP 
tau-functions which are infinite hypergeometric series of Schur 
polynomials (\ref{vh}),(\ref{hZ}). We shall use the fermionic 
representations of tau-functions \cite{JM}.

\section{ Hypergeometric tau-functions }
\subsection{ Additional symmetries and tau-function}
Let $r$ be a function of one variable.  We shall assume that  
$r(n)$ is finite for $n\in Z$.
Let $D=z\frac{d}{dz}$ acts on the space of functions $\{ z^n ;n \in Z \}$ . 
Then $r(D)z^n =r(n)z^n$. 
All functions of operator $D$ which we consider below are 
given via their
eigenvalues on this basis. 

Let us consider an abelian subalgebra in ${\widehat {gl}}(\infty)$ formed
by the following set of fermionic operators
\begin{equation}\label{dopsim}
A_k=\frac{1}{2\pi i}
\oint \psi^*(z) \left(\frac{1}{z}r(D)\right)^k \psi(z),
\quad k=1,2,\dots  ,
\end{equation}
where the operator $r(D)$ acts on all functions of $z$ from the 
right hand side.
For the collection of independent variables 
$\beta=(\beta_1,\beta_2,\dots)$ we denote
\begin{equation}\label{Abeta}
A(\beta)=\sum_{n=1}^\infty \beta_nA_n .
\end{equation}
For the partition ${\bf n}=(n_1,\dots ,n_k)$ and a function of one variable 
$r$, let us introduce the following notation
\begin{equation}\label{r_n}
r_{{\bf n}}(M)=\prod_{i=1}^{k}
r(1-i+M)r(2-i+M)\cdots r(n_i-i+M).
\end{equation}
We set $r_{{\bf 0}}(M)=1$.
Using the notation from (\ref{glemma}) we have\\
{\bf Lemma 2} The following formula holds
\be
\l 0|\psi^*_{i_1}\cdots\psi^*_{i_s}\psi_{-j_s}
\cdots\psi_{-j_1}e^{-A(\beta )}  |0\r
 =(-1)^{j_1+\dots +j_s} 
r_{{\bf n}}(0)s_{{\bf n}}(\beta).
\ee

Let us consider the following tau-function of the KP hierarchy
\be
\tau_r(M,{\bf t},\beta ):=\langle M|e^{H({\bf t})}
e^{-A({\bf \beta })}  |M\rangle .
\ee
{\bf Proposition 1} We have the following expansion:
\be\label{tauhyp}
\tau_r(M,{\bf t},\beta )
= \sum_{n=0}^{+\infty}\sum_{{\bf |n|}=n}
r_{ \bf n }(M)s_{\bf n }({\bf t})s_{\bf n }({\bf \beta }) .
\ee

We shall not consider the problem of convergence of this series. 
The variables $M,{\bf t}$ play the role of KP higher
times, $\beta$ is a collection of group times for a commuting
subalgebra of additional symmetries of KP (see \cite{OW,D',ASM} and 
Remark 7 in \cite{O}). From different point of view (\ref{tauhyp}) is a tau-function of two-dimensional Toda lattice \cite{UT} with two sets of
continuous variables ${\bf t}$, $\beta$ and one discrete variable $M$. 
Formula (\ref{tauhyp}) is symmetric with respect to   
${\bf t} \leftrightarrow\beta$.
This 'duality' supplies us with the string equation \cite{T} which 
characterizes a tau-function of hypergeometric type (see below).
In \cite{Tinit} the similar expansions to (\ref{tauhyp}) were considered,
without specifying the coefficients and in a different context.\\

Let us introduce 
\begin{equation}\label{tdopsim}
\tilde{A}_k=-\frac{1}{2\pi i}
\oint \psi^*(z) \left(\tilde{r}(D)z\right)^k \psi(z),
\quad (k=1,2,\dots),\quad
\tilde{A}(\beta)=\sum_{n=1}^\infty \tilde{\beta}_n\tilde{A}_n .
\end{equation}
Then we have the following generalization of Proposition 1:\\
{\bf Proposition 2}
\begin{equation}\label{trr}
\langle M|e^{\tilde{A}( {\tilde {\beta }})}
e^{-A({\bf \beta })}  |M\rangle
= \sum_{\bf n}
(\tilde{r}r)_{ \bf n }(M)s_{\bf n }(\tilde{\beta })s_{\bf n }({\bf \beta }).
\end{equation}

For an interpretation of this expansion see {\em Remark 2} below.
Let us mark that the function $r(M)$
is connected with the Sato $b$-function \cite{KV}.

Let us consider the following difference equations on functions 
$\tilde{h}(D),h(D)$:
\be\label{h}
\tilde{h}(D)\tilde{r}(D)\tilde{h}^{-1}(D-1)=1,\quad
h^{-1}(D-1)r(D)h(D)=1 .
\ee
The similar equation appeared in the paper of Graev
(see formula (6) in \cite{Graev}) and was used for generating 
of different hypergeometric series. As in \cite{Graev}, in terms
of the operator $r(D)$, it is possible to construct the differential
equations (and q-difference equations) for hypergeometric 
functions. We shall present some examples below in (\ref{ord-eq}) 
and (\ref{q-eq}).

Our fermionic representation 
is equivalent to that used in Toda lattice theory 
\cite{UT} if $r(n)\neq 0, n \in Z$.
We define a Hamiltonian $H_0(h)\in \widehat{gl}(\infty)$ and a set of
$C_n,n \in Z$:
\be\label{H0}
H_0(h):=\frac{1}{2\pi i}\oint :\psi^*(z) \log (h(D)) \psi(z):
\ee
\be
C_n=\frac{1}{h(n-1)}\cdots \frac{1}{h(1)}\frac{1}{h(0)}
\frac{1}{\tilde{h}(n-1)}\cdots \frac{1}{\tilde{h}(1)}\frac{1}{\tilde{h}(0)}, 
\quad n>0,
\ee
\be
C_n =h(n)\cdots h(-2)h(-1)
\tilde{h}(n)\cdots\tilde{h}(-2)\tilde{h}(-1),\quad n<0 .
\ee
{\bf Proposition 3} If function $r$ has no zeroes at integer
values of argument then
\begin{eqnarray}\label{teplitsGauss}
\tau(n,{\tilde{\beta}},\beta):=
\langle n|e^{\tilde{A}({\tilde{\beta}})}e^{-A({\beta})}|n \rangle =
\langle n|e^{H({\tilde{\beta}})}g e^{-H^*({\beta})}|n \rangle C^{-1}_n ,\\
g=e^{H_0(\tilde{h}h)}=e^{H_0(\tilde{h})+H_0(h)},\quad
C_n=\langle n| g |n \rangle . \label{g}
\end{eqnarray}

For $\tilde{r}=1$ we can put $\tilde{\beta}={\bf t}$. Then
the following equations hold
\begin{eqnarray}\label{rToda}
\partial_{t_1}\partial_{\beta_1}\phi_n=
r(n)e^{\phi_{n-1}-\phi_{n}}-
r(n+1)e^{\phi_{n}-\phi_{n+1}} ,\quad
e^{-\phi_n}=\frac{\tau(n+1,{\bf t},\beta)}{\tau(n,{\bf t},\beta)},\\
\tau(n)\partial_{\beta_1}
\partial_{t_1}\tau(n)-
\partial_{t_1}\tau(n)
\partial_{\beta_1}\tau(n)=r(n)\tau(n-1)\tau(n+1).\label{hirota}
\end{eqnarray}

If the function $r$ has no integer zeroes, then after the change $\varphi_n=-\phi_n - \log h(n)$  we obtain Toda lattice equation in standard form \cite{UT}:
\be\label{Toda}
\partial_{t_1}\partial_{\beta_1}\varphi_n=e^{\varphi_{n+1}-\varphi_{n}}-
e^{\varphi_{n}-\varphi_{n-1}}.
\ee

The main point of the paper is based on
the observation that if $r(D)$ is a rational function
of $D$ then $\tau_r$ is a hypergeometric series. If  $r(D)$ is
a rational function of $q^D$ we obtain q-deformed hypergeometric series.
 We shall see both cases in the following examples.
(In a separate paper the case of rational expressions of elliptic
theta-functions will be considered).

\subsection{Examples of the tau-functions}
Now let us consider various $r(D)$.\\
{\bf Example 1}
Let $r(M)=M$ and
 $\beta=(\beta_1,0,0,...)$. 
 For $M=\pm1$ we get 
\be
\tau(1,{\bf t},\beta_1)=e^{\xi({\bf t},\beta_1)},\quad
\tau(-1,{\bf t},\beta_1)=e^{-\xi({\bf t},\beta_1)}.
\ee
Thus $\beta_1$ plays the role of a spectral parameter for the vacuum
Baker-Akhiezer function. 
This is in accordance to the meaning of $\beta_1$ as a group time for the 
Galilean transformation \cite{O}.\\
{\bf Example 2} Let all parameters $b_k$ be nonintegers.
\be
{}_pr_s(D)=\frac{(D+a_1)(D+a_2)\cdots 
(D+a_p)}{(D+b_1)(D+b_2)\cdots (D+b_s)} . \label{op1}
\ee
If all $a_k$ are also nonintegers the
relevant $h(D)$ is:
\be
{_ph_s}(D)=\frac{\Gamma(D+b_1+1)\Gamma(D+b_2+1)\cdots 
\Gamma(D+b_s+1)}{\Gamma(D+a_1+1)\Gamma(D+a_2+1)\cdots 
\Gamma(D+a_p+1)} . \label{op2}
\ee
For the correlator (\ref{tauhyp}) we have:
\be
{}_p\tau_s(M,{\bf t},\beta)=
\sum_{n=0}^{+\infty}\sum_{|{\bf n}|=n}s_{{\bf n}}({\bf t})s_{{\bf n}}(\beta)
\frac{(a_1+M)_{{\bf n}}\cdots (a_p+M)_{{\bf n}}}
{(b_1+M)_{{\bf n}}\cdots (b_s+M)_{{\bf n}}} . \label{tau}
\ee

One can get the hypergeometric function related to 
Schur functions \cite{KV} by putting $\beta_1=1$ and $\beta_i=0$ for 
$i=(2,3,\dots)$ in (\ref{tau}):
\begin{eqnarray}\label{beskshur}
{}_pF_s\left.\left(a_1+M,\cdots ,a_p+M\atop b_1+M,\cdots ,b_s+M\right|t_1,t_2,\dots\right)=
\nonumber\\
\sum_{n=0}^{+\infty}
\sum_{{\bf |n|}=n}\frac{(a_1+M)_{{\bf n}}\cdots (a_p+M)_{{\bf n}}}
{(b_1+M)_{{\bf n}}\cdots (b_s+M)_{{\bf n}}}
\frac{s_{{\bf n}}({\bf t})}{H_{{\bf n}}}\label{ss}
\end{eqnarray}
with hook polynomial $H_{{\bf n}}=s_{{\bf n}}({\beta})=\prod_{(i,j)\in {\bf n}} h_{ij}$, $h_{ij}=(n_i+n'_j-i-j+1)$ \cite{Mac}.

We obtain the ordinary hypergeometric function of one variable of type 
\be\label{onevar}
{}_{p-1}F_s(\pm t_1\beta_1)=\tau(\pm 1,{\bf t},\beta ) ,
\ee
 if we take $a_1=0$, ${\bf t}=(t_1,0,0,\dots)$, 
$\beta =(\beta_1,0,0,\dots)$ \cite{Sc}.

Now we consider the following change of variables
$t_m=\sum_{i=1}^N \frac {x_i^m}{m}$.
In this case the formula (\ref{beskshur}) turns out to be
the hypergeometric function of matrix argument, see (\ref{hZ}).
Taking $N=1$ we obtain the ordinary hypergeometric function of
one variable, $x_1$. Formula (\ref{sym=miwa}) will explain
 the connection
between this hypergeometric function and the function (\ref{onevar}).
The ordinary hypergeometric function satisfies well-known hypergeometric 
equation
\be\label{ord-eq}
\left(\partial_{x_1}-{}_pr_s(D_1)\right)
{}_pF_s(x_1)=0,
\quad D_1:=x_1\partial_{x_1} .
\ee
\\
{\bf Example 3} 
The q-generalization of the {\em Example 2}:
\be\label{qr}
{}_pr_s^{(q)}(D)= 
\frac{\prod_{i=1}^p (1-q^{a_i+D})}{\prod_{i=1}^s(1-q^{b_i+D})} .
\ee
For the variables  $\beta_i=\frac{1}{i(1-q^i)}$ ($i=1, 2,\dots$) 
and $t_m=\sum_{j=1}^N \frac{x_j^m}{m}$ we obtain the 
formula (\ref{vh}) (see \cite{Mac})
\begin{eqnarray}
{}_p\tau_s^{(q)}\left.\left(a_1+M,\dots ,a_p+M\atop b_1+M,\dots ,
b_s+M\right|q,{\bf x}_{(N)}\right)=\nonumber\\
\sum_{n=0}^{+\infty}\sum_{|{\bf n}|=n\atop l({\bf n})\le N }
\frac{(q^{a_1+M}; q)_{{\bf n}}\cdots (q^{a_p+M};q)_{{\bf n}}}
{(q^{b_1+M};q)_{{\bf n}}\cdots (q^{b_s+M};q)_{{\bf n}}}
\frac{q^{n({\bf n})}}{H_{{\bf n}}(q)}s_{{\bf n}}({\bf x}_{(N)}) .
\label{qtau}
 \end{eqnarray}
When $N=1$ we have the ordinary q-deformed hypergeometric function:
\begin{eqnarray}\label{qss1}
{}_p\Phi_s\left.\left(a_1+M,\cdots ,a_p+M\atop b_1+M,\cdots ,b_s+M\right|q,x_1\right)
=\sum_{n=0}^{+\infty}\frac{(q^{a_1+M};q)_{n}
\cdots (q^{a_p+M};q)_{n}}{(q^{b_1+M};q)_{n}\cdots (q^{b_s+M};q)_{n}}
\frac{x_1^{n}}{(q;q)_n}
\end{eqnarray}
which satisfies the following q-difference equation
\be\label{q-eq}
\left(\frac {1}{x_1} \left(1-q^{D_1}\right)-{}_p r_s^{(q)}(D_1)\right)
{}_p\Phi_s(x_1)=0,
\quad D_1:=x_1\partial_{x_1} .
\ee
There are various applications for series (\ref{qss1}), for instance see
\cite{QCD},\cite{Pugai} and \cite{Odz}. Bosonic representation of 
(\ref{qss1}) was found in \cite{MV}.
 Let us note that operator $q^D$ which acts on fermions $\psi(z)$
was used in \cite{LS} in different context.\\
{\bf Example 4} q-{\bf Askey-Wilson polynomials}\\
Let operator ${}_4r^{(q)}_3(D)$ be
\be\label{rq-askey}
{}_4r_3^{(q)}(D)= \frac{(1-q^{-n+D})(1-abcdq^{n-1+D})(1-ae^{i\eta}q^D)(1-ae^{-i\eta}q^D)}{(1-abq^{D})(1-acq^D)(1-adq^D)} .
\ee
By choosing  $\beta_i=\frac{1}{i(1-q^i)}$ and 
${\bf t}=(q,\frac{q^2}{2},\frac{q^3}{3},\dots)$ we get
\begin{eqnarray}\label{q-askey}
{}_4\tau_3^{(q)}(M,{\bf t},\beta)={}_4\varphi_3\left.\left(q^{M-n}, q^{M+n-1}abcd, aq^Me^{i\eta}, aq^Me^{-i\eta}\atop q^Mab,\quad q^Mac,\quad q^Mad\right| q,q\right)
=\nonumber\\ \sum_{m=0}^{+\infty}
\frac{(q^{M-n};q)_{m}(q^{M+n-1}abcd;q)_m(aq^Me^{i\eta};q)_m
(aqe^{-i\eta};q)_m}{(abq^{M};q)_{m}(acq^M;q)_m (adq^{M};q)_{m}}
\frac{q^{m}}{(q;q)_m} .
\end{eqnarray}
For $M=0$ we obtain q-Askey-Wilson polynomials:
\begin{eqnarray}
p_{n}(\cos\eta;a,b,c,d|q)=
aq^{-n}(ab;q)_{n}(ac;q)_{n}(ad;q)_{n} {}_4\tau^{(q)}_3({\bf t},\beta, 0) .
\end{eqnarray}
\\
{\bf Example 5} {\bf Clebsch-Gordan coefficients} 
$C_q({\bf l},{\bf j})$ see \cite{V}.\\
Let us take
\be
{}_3r_2^{(q)}(D)= \frac{(1-q^{j-l_1+D})(1-q^{l_1+j+1+D})(1-q^{-l+m+D})}
{(1-q^{l_2-l+j+1+D})(1-q^{-l-l_2+j+D})} .
\ee
For the variables ${\bf t}=(q, \frac{q^2}{2},\frac{q^3}{3},\dots)$ and
$\beta_i=\frac{1}{i(1-q^i)}$
\be
{}_3\tau^{(q)}_2(0,{\bf t},\beta)=
{}_3\Phi_2\left.\left(j-l_1, l_1+j+1, -l+m
\atop l_2-l+j+1,-l-l_2+j\right|q,q\right) .
\ee
Thus we have the following fermionic representation: $C_q({\bf l},{\bf j})=$
\be
\frac{(-1)^{l_1-j}q^B\Delta({\bf l})[l+l_2-j]!([{\bf l},
{\bf j}][2l+1])^{\frac{1}{2}}}{[l_1-l_2+l]!
[l+l_2-l_1]![l_2-l+j]![l_1-j]![l_2+k]![l-m]!} 
{}_3\tau^{(q)}_2(0,{\bf t},\beta),
\ee
\[
\left[a\right]:=
q^{(1-a)/2}\frac{1-q^a}{1-q},\quad [n]!:=[1][2]\cdots [n],\quad m=j+k, 
\]
\[
\Delta({\bf l})\equiv\Delta(l_1,l_2,l):=
\left(\frac{[l_1+l_2-l]![l_1-l_2+l]![l-l_1+l_2]!}{[l_1+l_2+l+1]!}
\right)^{\frac{1}{2}},
\]
\[
\left[{\bf l},{\bf j}\right]=[l_1+j]![l_1-j]![l_2+k]![l_2-k]![l+m]![l-m]!,
\]
\[
B=\frac{1}{4}(l_2(l_2+1)-l_1(l_1+1)-l(l+1)+2j(m+1)).
\]
\subsection{Different representations}
Let us rewrite  hypergeometric series in different way representing 
all Pochhammer coefficients $(q^a;q)_{\bf n}$ and $(a)_{\bf n}$  
 through Schur functions.  This gives us the opportunity to
interchange the role of Pochhammer coefficients and Schur
functions in (\ref{vh2}),(\ref{beskshur}), and to present different 
fermionic representations of the hypergeometric functions.
We have the following relations (see \cite{Mac}):
\be\label{PochSchur}
\prod_{(i,j)\in {\bf n}} (1-q^{a+j-i})=
\frac{s_{{\bf n}}({\bf t}(a,q))}{s_{{\bf n}}({\bf t}(+\infty,q))},\quad 
\prod_{(i,j)\in {\bf n}}(a+j-i)=\frac{s_{{\bf n}}({\bf t}(a))}{s_{{\bf n}}({\bf t}(+\infty)} ,
\ee
where parameters $t_m(a,q)$ 
and $t_m(a)$ are chosen via generalized Miwa transform \cite{Miwa} 
with multiplicity $a$  
\begin{eqnarray}\label{t(a)}
t_m(a,q)=\frac{1-(q^a)^{m}}{m(1-q^m)},\quad  
t_m(a)=\frac{a}{m},
\quad m=1, 2,\dots ,\\
 s_{{\bf n}}({\bf t}(+\infty,q))=\lim_{a\to +\infty}
s_{{\bf n}}({\bf t}(a,q))=\frac{q^{n({\bf n})}}{H_{{\bf n}}(q)} ,\\
 s_{{\bf n}}({\bf t}(+\infty))=\lim_{a\to +\infty}
s_{{\bf n}}\left(\frac{t_1(a)}{a}, \frac{t_2(a)}{a^2},\dots\right)=
\lim_{a\to +\infty}\frac{1}{a^{|{\bf n}|}}s_{{\bf n}}({\bf t}(a))=
\frac{1}{H_{{\bf n}}} .
\end{eqnarray}
Now we rewrite the series (\ref{qtau}) and (\ref{beskshur}) only in terms
of Schur functions:
\begin{eqnarray}
{}_p\Phi_s\left.\left(a_1+M,\dots ,a_p+M\atop b_1+M,
\dots ,b_s+M\right|q,{\bf x}_{(N)}\right)=
\tau_{r}(M,{\bf t}(+\infty,q),{\bf t}) \nonumber\\ 
=\sum_{n=0}^{+\infty}\sum_{|{\bf n}|=n\atop l({\bf n})\le N }
\frac{\prod_{k=1}^p{s_{{\bf n}}({\bf t}(a_k+M,q))}}
{\prod_{k=1}^s{s_{{\bf n}}({\bf t}(b_k+M,q))}}
\left(s_{{\bf n}}({\bf t}(+\infty,q))\right)^{s-p+1}
s_{{\bf n}}({\bf x}_{(N)}) , \label{qtaushur}
\end{eqnarray}
\begin{eqnarray}
{}_pF_s\left.\left(a_1+M,\cdots ,a_p+M\atop b_1+M,\label{taushur}
\cdots ,b_s+M\right| {\bf x}_{(N)}\right)=
\tau_{r}(M,{\bf t}(+\infty),{\bf t})=\nonumber\\
\sum_{n=0}^{+\infty}\sum_{|{\bf n}|=n\atop l({\bf n})\le N }
\frac{\prod_{k=1}^p s_{{\bf n}}({\bf t}(a_k+M))}
{\prod_{k=1}^s s_{{\bf n}}({\bf t}(b_k+M))}
\left(s_{{\bf n}}({\bf t}(+\infty))\right)^{s-p+1}s_{{\bf n}}({\bf x}_{(N)}).
\end{eqnarray}
We obtain different fermionic representations of hypergeometric 
functions (\ref{taushur}), (\ref{qtaushur}) and they are parametrized
by a complex number $b$:\\
{\bf Proposition 4}. For $b \in C$ and for $r= {}_pr_s$ 
(see (\ref{op1})) we have
\be\label{sym=miwa}
\tau_r(M,{\bf t}(+\infty),{\bf t})=\tau_{r_b}(M,{\bf t}(b+M),{\bf t}),
\quad r_b=\frac{r}{b+D}.
\ee
For $r= {}_pr_s^{(q)}$ (see (\ref{qr})) we have
\be\label{qsym=miwa}
\tau_r(M,{\bf t}(+\infty,q),{\bf t})=\tau_{r_b}(M,{\bf t}(b+M,q),{\bf t}),
\quad r_b=\frac{r}{1-q^{b+D}} .
\ee
{\bf Remark 1}. There are two ways to restrict the sum 
(\ref{tauhyp}) to the sum over partitions of length $l({\bf n})\leq N$.
 First, if we use Miwa's change (\ref{tmiwa}), then
$s_{{\bf n}}({\bf x}_{(N)})=0$, for ${\bf n}$ with length $l({\bf n})>N$.
The second way is to restrict the  Pochhammer coefficients:
if we put  $a_i=N$ for one $i$ from (\ref{qr}) equal to $N$, then the 
coefficient (\ref{PSimb}) vanishes for  $l({\bf n})>N$.
Since we expressed Pochhammer coefficients in terms of Schur functions
in (\ref{PochSchur}) both ways have the same explanation.
Indeed
\be
t_m(N,q)=\frac{1}{m}\frac{1-(q^N)^m}{1-q^m}=
\frac{1}{m}(1+(q)^m+(q^2)^m+\dots+(q^{N-1})^m) .
\ee
Therefore we obtain for Miwa's change: $x_1=1,x_2=q,\dots , x_N=q^{N-1}$ and
\be
s_{{\bf n}}({\bf t}(N,q))=s_{{\bf n}}(1, q,\dots, q^{N-1})=0,
\quad l({\bf n})>N.
\ee
The same we have for the sum
over partitions ${\bf n}$ such that $l({\bf n'})<K$.
Again the first way has to be realized through the following Miwa's 
change of variables:
\be\label{mtmiwa}
t_m=-\sum_{i=1}^{K}\frac{x_i^m}{m},\quad s_{{\bf n}}({\bf t})=s_{{\bf n'}}({\bf x}_{(K)}) .
\ee
The second way is to make one of the parameters, for example  $a_j$ 
from (\ref{qr}) equal to $(-K)$. In this case
\be
s_{{\bf n}}({\bf t}(-K,q))=s_{{\bf n'}}\left(\frac{1}{q},\frac{1}{q^2},\dots,\frac{1}{q^{K}}\right)=0, \quad l({\bf n'})>K.
\ee

\subsection{Zakharov-Shabat dressing and string equations}
Now we shall get hypergeometric functions from 
another point of view - the Zakharov-Shabat factorization problem.
Let us introduce infinite matrices to describe KP and Toda lattice
flows and symmetries \cite{UT}. 
We denote the Zakharov-Shabat dressing 
matrices by $K$ and $\bar{K}$. $K$ is a lower triangular matrix with unit 
main diagonal, and $\bar{K}$ is an upper triangular matrix.
They solve Gauss factorization problem for infinite 
matrices: 
\be\label{faktorKP}
\bar{K}=KU(M,{\bf t},\beta),\quad
U(M,{\bf t},\beta)=U^{+}({\bf t})U^{-}(M,\beta ).
\ee
Here $U^{\pm}$ belong to the different abelian multiparametric subgroups in 
$GL(\infty)$ with the infinite sets of group times ${\bf t}$ 
and $\beta$,
\be
U^{+}({\bf t})=\exp \left( \xi({\bf t},\Lambda)\right),\quad
U^{-}(M,\beta )=
\exp \left(\xi(\beta,\Lambda ^{-1}r\left(\Delta+M\right))\right),
\ee
where the matrices $(\Lambda)_{jk}=\delta_{j,k-1}$,
$(\Delta)_{jk}=j\delta_{j,k}$, for $\xi$ see (\ref{KPvar}). 
The function $r$ is the same as 
in (\ref{dopsim}).
Then following Zakharov-Shabat arguments we find that 
the variables $-\log (\bar{K}_{ii})=\phi_{i+M}$ solve Toda lattice equation 
in the form (\ref{rToda}). At the same time (\ref{faktorKP})
describes a set of KP equations \cite{UT} parametrized by integer $M$.

The tau-function can be obtained as follows.
By taking the projection \cite{UT} $U \mapsto U_{--}$ for nonpositive 
values of matrix 
indices we obtain a determinant representation of 
the tau-function (\ref{tauhyp}):
\be \label{2-c}
\tau_r(M,{\bf t},\beta)=
\frac {\det U_{--}(M,{\bf t},\beta)}
{ \det \left(U^{+}_{--}({\bf t}) \right)
\det \left(U^{-}_{--}(M,\beta ) \right)}
=\det U_{--}(M,{\bf t},\beta) ,
\ee
since both determinants in the denominator are equal to one.
Formula (\ref{2-c}) is also a Segal-Wilson formula for ${GL}(\infty)$ 
2-cocycle \cite{SW}. Choosing the function $r$ as in {\em Section 3.2}
we obtain hypergeometric functions listed in the {\em Introduction}.\\
{\bf Remark 2}. Therefore the hypergeometric functions which were 
considered above
have the meaning of $GL(\infty)$ two-cocycle on the two 
multiparametrical group elements $U^{+}({\bf t})$ and  
$U^{-}({M,\beta})$. 
Both elements $U^{+}({\bf t})$ and  $U^{-}(M,\beta )$  can be considered as
elements of group
of pseudodifferential operators on the circle. The corresponding Lie 
algebras consist of the multiplication operators $\{z^n;n \in N_0\}$
and of the pseudodifferential operators
$\{\left( \frac 1z r(z\frac {d}{d z}+M) \right)^n;n \in N_0\}$.
Two sets of group 
times ${\bf t}$
and $\beta$ play the role
of indeterminates of the hypergeometric functions (\ref{tau}).
Formulas (\ref{tauhyp}) and (\ref{trr}) mean the expansion
of ${GL}(\infty)$ group 2-cocycle in terms of corresponding
Lie algebra 2-cocycle
\be
\omega (z,\frac 1z r(D+M))=r(M),\quad 
\omega (\tilde{r}(D+M)z,\frac 1z r(D+M))=\tilde{r}(M)r(M).
\ee

Now following \cite{TT,T,NTT} let us briefly describe additional symmetries  mentioned in {\em Section 3.1} and 
the string equations. Below we use the notations from the papers
\cite{T,NTT} for $L,\bar{L},\widehat{M};
[\widehat{M},L]=1$.
The Toda lattice additional symmetries and higher
flows are given by the following Lax equations
\be
\partial_{\beta_n}L=[L,
\left(\left(r(\widehat{M})L^{-1}\right)^n\right)_- ],\quad
\partial_{\beta_n}\bar{L}=[\bar{L},
\left(\left(r(\widehat{M})L^{-1}\right)^n\right)_+],\quad
\ee
\be
\partial_{t^*_n}L=[L,\left(\bar{L}^n\right)_-],\quad
\partial_{t^*_n}\bar{L}=[\bar{L},\left(\bar{L}^n\right)_+].\
\ee
Let us impose the condition that the group times $\beta_m$ of
additional symmetries can be 
identified with the Toda lattice times $t^*_m$.
Then one can obtain a set of string equations \cite{TT,T},
which characterizes the TL hypergeometric solutions:
\be
h(\widehat{M}-1) L^m=\bar{L}^m h(\widehat{M}-1),\quad m=2,3,\dots ,
\ee
where functions $h$ and $r$ are connected by (\ref{h}).
The simplest string equation is $L=\bar{L} r(\widehat{M})$.
When $r(M)=M+a$ the string equation describes $c=1$ string \cite{NTT,T}. 

At last let us mention the $\bar{\partial}$ problem \cite{ZMan} 
for the KP-1 equation
\be\label{dbar}
\frac{\partial w(M,{\bf t},{\bf t}^*,z,\bar{z})}{\partial \bar{z}}=
w(M,{\bf t},{\bf t}^*,\bar{z},z)
T(z,\bar{z}) ,
\ee
where for $z\to \infty$ and for $z\to 0$ the asymptotics of Baker
function $w$ are
\be
w=z^Me^{\xi({\bf t},z)}
\left(1+O(z^{-1})\right),\quad 
w=z^{-M}e^{-\xi({\bf t}^*,z^{-1})}
e^{\varphi_M({\bf t},{\bf t}^*)}
\left(1+O(z)\right)
\ee
with some function $\varphi_M$.
Instead of writing down
the explicit form of the function $T$, which gives the hypergeometric
solution of the KP-1 equation, we only give a set of constraints on $T$
(which are equivalent to the conditions mentioned above: ${\bf t}^*$
may be identified with times of additional symmetries $\beta$ \cite{O}):
\be\label{symT} 
\left(\frac1zr(D)\right)^mT(z,\bar{z})=
\left(\frac{1}{\bar{z}}r(-\bar{D})\right)^m
T(z,\bar{z}),\quad m=1,2,\dots .
\ee
For the KP-2 equation we have nonlocal Riemann problem, see \cite{ZMan}
\be\label{nRp}
w_+(z)-w_-(z)=\int_R
w_-(z')
R(z',z)dz'
\ee 
with the similar constraints on the kernel $R$:
\be\label{symR} 
\left(\frac1zr(z\partial_z)\right)^mR(z',z)=
\left(\frac{1}{z'}r(-z'\partial_{z'})\right)^m
R(z',z),\quad m=1,2,\dots .
\ee
The statement is the following. If a hypergeometrical
solution belongs to the class described by (\ref{dbar}) (or by (\ref{nRp})),
then the function $T$ (respectively $R$) solves  equation (\ref{symT})
 (respectively (\ref{symR})). 
 
We present the infinitesimal version of the Zakharov-Shabat 
dressing \cite{ZSh}
\be
\frac{1}{2\pi i}\oint \left(\left(z^*-\frac{1}{z}r(D)\right)^{-1}w(z)\right)
\partial_x^{-1} w^*(z)dz=\sum_{n=1} {z^*}^{-n-1}{\itshape{A}_n} 
\ee
as a generating function for the Zakharov-Shabat zero curvature equations
for the additional symmetries $[\partial_{t_n}-B_n,
\partial_{\beta_m} -{\itshape{A}_m}]=0$, 
compare with \cite{ASM,D',O}. \\

From soliton theory and bosonization formula one
can obtain various relations for the tau-function (\ref{tauhyp}).
Let us consider an example. We introduce
\be
\Omega_r(\beta):=
-\frac{1}{2\pi i}\lim_{\epsilon \to 0} \oint 
 V^*(z+\epsilon)\sum_{m=1}^\infty \beta_m 
\left(\frac{1}{z}r(z,D)\right)^m
V(z)dz ,
\ee
where $V(z)$ and $V^*(z)$ are defined by (\ref{vertex}). 
Now we have the following shift argument formula
\be
e^{\Omega_r(\gamma)}\tau_r(M,{\bf t},\beta)=\tau_r(M,{\bf t},\beta +\gamma).
\ee

\subsection{Orthogonal polynomials}
It is known that the hypergeometric functions (\ref{hZ}) appear 
in the group representation theory and are connected with 
the so-called matrix integrals \cite{V}. On the other hand
the set of examples \cite{GMMO,Ko} reveals a
connection between the matrix integrals and the soliton theory.
To establish this connection it is useful to consider the related
systems of the orthogonal polynomials. Let us briefly describe how to
write down these polynomials.
Let $M_+$ be the largest integer zero of the function $r$. 
Then the function
\be
f(zz^*)=\sum_{n=0}^{+\infty}(zz^*)^{n+M_+} h(n+M_+)
\ee
is the eigenfunction of the operator $\frac{1}{z}r(D)$ 
with the eigenvalue $z^*$. We use this function as weight function for
a system of orthogonal polynomials $p_n^{\pm},n=0,1,2,\dots$,
related to the hypergeometric solution of KP:
\be
\int_\gamma p_n^{-}(z,{\bf t},\beta)
e^{\xi(z,{\bf t})+\xi(z^*,{\beta})}f(zz^*) 
p_m^{+}(z^*,{\bf t},\beta)dzdz^*
=e^{-\phi_{M_{+}+ n}({\bf t},\beta)}\delta_{n,m} .
\ee
\subsection{Further generalization}
Formula (\ref{teplitsGauss}) is related to 'Gauss decomposition'
of operators inside vacuums $\langle 0|\dots|0\rangle $ into
diagonal operator $e^{H_0(h)}$, upper triangular operator $e^{H({\bf t})}$  and lower triangular operator $e^{-H^*({\bf t}^*)}$ (the last two have the Toeplitz form). Now let us consider more general two-dimensional 
Toda lattice tau-function
\begin{eqnarray}\label{general}
\tau=\l M|e^{H({\bf t})}ge^{-A(\beta)}|M\r,\quad
g=e^{\tilde{A}_1(\tilde{\gamma}_1)}
\cdots e^{\tilde{A}_k(\tilde{\gamma}_k)}e^{-A_l(\gamma_l)}
\cdots e^{-A_1(\gamma_1)} ,
\end{eqnarray}
where operators $A_i(\g_i)=\sum_{j=1}^{+\infty}\g_{ij}A_{ij}$ and $\tilde{A_i}(\tilde{\g}_i)=\sum_{j=1}^{+\infty}\tilde{\g}_{ij}\tilde{A}_{ij}$ are defined like $A(\beta)$ of (\ref{dopsim}) and $\tilde{A}(\tilde{\beta})$ of (\ref{tdopsim}) respectively and correspond to operators $r^i(D)$ and $\tilde{r}^i(D)$.
 Collections of variables $\tilde{\gamma}=\{\tilde{\gamma}_{ij}\},
\gamma=\{\gamma_{ij}\}$ play the role of coordinates for some wide enough 
class of Clifford group elements $g$.
Let us calculate this tau-function. We introduce a set consisting of $m+1$ partitions:
\be
({\bf n_1},\dots ,{\bf n_m}, {\bf n_{m+1}}={\bf n}),\quad 
0\le {\bf n_1}\le {\bf n_2}\le \cdots\le {\bf n_m}\le {\bf n_{m+1}}={\bf n}.
\ee
  Also we define a set $\Theta_{{\bf n}}^m=({\bf n_1},{\bf n_2}/{\bf n_1},\dots ,{\bf n}/{\bf n_m})$.  We take    
\be
s_{\Theta^m_{{\bf n}}}({\bf \mu})=s_{{\bf n_1}}(\mu_1)
s_{{\bf n_2}/{\bf n_1}}(\mu_2)\cdots s_{{\bf n}/{\bf n_m}}(\mu_{m+1}),\quad
\mu_{i}=\{\mu_{ij}\} .
\ee
Here $s_{{\bf n_{i+1}}/{\bf n_i}}(\mu_{i+1})$ is a skew Schur function \cite{Mac}:
\be
s_{{\bf n_{i+1}}/{\bf n_i}}(\mu_{i+1})=\det(p_{n^{i+1}_{\rho}-n^i_{\nu}-\rho+\nu}(\mu_{i+1}))_{1\le \rho,\nu\le r},\quad {\bf n_{i+1}}=(n^{(i+1)}_{1},\dots ,n^{(i+1)}_{r}).
\ee

A skew analogy of  (\ref{r_n}) is
\begin{eqnarray}
r_{\Theta^m_{{\bf n}}}(M)=r_{{\bf n_1}}(M)r^{1}_{{\bf n_2}/
{\bf n_1}}(M)\cdots r^{m}_{{\bf n}/{\bf n_m}}(M) , \\
r_{{\bf n_{i+1}}/{\bf n_i}}(M)=\prod_{j=1}^r r(n^{(i)}_{j}-j+1+M)
\cdots r(n^{(i+1)}_{j}-j+M).
\end{eqnarray}
If the function $r^i(m)$  has no poles and zeroes at integer points, then the relation $r^{i}_{\theta_i}(M)=\frac{r^{i}_{{\bf n_{i+1}}}(M)}{r^{i}_{{\bf n_{i}}}(M)}$
($i=1,\dots , m$) is correct. To calculate the tau function we need the following
{\em Lemma}\\
{\bf Lemma 3} Let partitions 
\be
{\bf n}=(i_1,\dots ,i_s|j_1-1,\dots ,j_s-1),\quad {\bf \tilde{n}}=(\tilde{i}_1,\dots ,\tilde{i}_r|\tilde{j}_1-1,\dots ,\tilde{j}_r-1)
\ee
 satisfy the relation ${\bf n}\ge{\bf \tilde{n}}$.
Then the next formula is valid:
\begin{eqnarray}
\l 0|\psi^*_{\tilde{i}_1}\cdots\psi^*_{\tilde{i}_r}\psi_{-\tilde{j}_r}\cdots
\psi_{-\tilde{j}_1}e^{A^i(\g_i)}\psi^*_{-j_1}\cdots
\psi^*_{-j_s}\psi_{i_s}\cdots\psi_{i_1}|0\r=\nonumber\\
=(-1)^{\tilde{j}_1+\dots+\tilde{j}_r+j_1+\dots+j_s}s_{\theta}(\g_i)r_{\theta}(0),\qquad \theta={\bf n}/\tilde{{\bf n}} .
\end{eqnarray}
The proof is achieved by direct calculation (see \cite{Mac}
for help).\\
Now  we obtain the following generalization of {\em Proposition 1}:\\
{\bf Proposition 5}
\be\label{gen}
\tau(M,{\bf t},\beta ;\g , \tilde{\g})=\sum_{{\bf n}}\sum_{\Theta^k_{{\bf n}}}
\sum_{\Theta^l_{{\bf n}}} \tilde{r}_{\Theta^k_{{\bf n}}}(M)r_{\Theta^l_{{\bf n}}}(M)
s_{\Theta^k_{{\bf n}}}({\bf t},\tilde{\g}) s_{\Theta^l_{{\bf n}}}(\beta,\g) .
\ee 
With the help of this series one can obtain different hypergeometric
series.\\
{\bf Example 6}. Let us consider the tau function given by the correlator
\begin{eqnarray}
\tau(M,\tilde{\beta},\beta,\g_1)=
\l M|e^{\tilde{A}(\tilde{\beta})}e^{-A_1(\g_1)}e^{-A(\beta)}|M\r , \\
\tilde{r}^1(D)=\frac{\tilde{a}_1+D}{\tilde{b}_1+D},\quad r(D)=\frac{a_1+D}{b_1+D},\quad r^1(D)=1 .
\end{eqnarray}
We put $\tilde{\beta}=(x, \frac{x^2}{2},\frac{x^3}{3},\dots)$, $\beta=(y_1,0,0,\dots )$, $\g_1=(y_2,0,\dots )$.
Thus we have 
\be
\tau(M, x, y_1, y_2)=\sum_{n_1=0}^{+\infty}\sum_{n_2=0}^{+\infty}
\frac{(\tilde{a}_1+M)_{n_1+n_2}(a_1+M)_{n_1}}{(\tilde{b}_1+M)_{n_1+n_2}(b_1+M)_{n_1}}
\frac{y_1^{n_1}y_2^{n_2}}{n_1!n_2!}x^{n_1+n_2} .
\ee

\section*{Conclusion}

We get multivariate hypergeometric functions as certain tau-functions
of the KP hierarchy and also as the ratios of Toda lattice tau-functions 
considered in \cite{Tinit}, \cite{T} evaluated at certain
values of higher Toda lattice times.
It means that multivariate  hypergeometric functions solve a set
of continuous and discrete bilinear Hirota equations \cite{JM}. We shall 
write down these equations explicitly in the different paper.
Hypergeometric solution of the KP equation is 
$u=2\partial^2_{t_1}\log \tau$,
where $\tau$ is a hypergeometric function.
To investigate the properties of this new class of solutions is
a separate interesting problem which we leave for future investigation.
It is quite unexpected that we get q-deformed version of these hypergeometric
functions as tau-functions not of a q-deformed KP hierarchy \cite{Bogd}
but of the 
usual KP hierarchy. It is now an
 interesting problem to establish links between these results,
 group-theoretic approach to the 
q-special functions \cite{V,KV} and matrix integrals. 
We expect to work out connections with matrix models of Kontsevich 
type \cite{Ko} and two-matrix models related to 2D Toda lattice 
\cite{GMMO}. We can present links between our construction
in the present paper and so-called generalized Miwa's change of variables
\cite{Miwa} in the
three dimensional three-wave systems \cite{ZSh}, and with the multicomponent
KP hierarchy \cite{JM}, it will be published 
separately.
We hope to generalize our results to the KP hierarchies
of $B_\infty$, $C_\infty$ and $D_\infty$ types \cite{JM} to get different 
hypergeometric series.

\section*{Acknowledgements}
One of the authors (A.O.) is pleased to thank T.Shiota and especially 
Vl.Dragovich for the helpful discussions, and also L.A.Dickey for the 
interest.  D.S. would like to thank S.Senchenko for helpful discussions.

\end{document}